\documentclass[aps,prd,showpacs]{revtex4}
\input epsf.sty
\begin{document} 
\hspace{-10mm} 
\leftline{\epsfbox{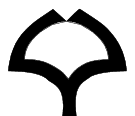}} 
\vspace*{-10.0mm} 
\thispagestyle{empty} 
{\baselineskip-4pt 
\font\yitp=cmmib10 scaled\magstep2 
\font\elevenmib=cmmib10 scaled\magstep1  \skewchar\elevenmib='177 
\leftline{\baselineskip20pt 
\hspace{10mm} 
\vbox to0pt 
   { {\yitp\hbox{Osaka \hspace{1.5mm} University} } 
     {\large\sl\hbox{{Theoretical Astrophysics}} }\vss}} 
 
\rightline{\large\baselineskip20pt\rm\vbox to20pt{ 
\baselineskip14pt 
\hbox{OU-TAP-194}
\vspace*{1mm} 
\hbox{\today}\vss}}
} 
\vskip8mm 
\begin{center}{\large \bf 
Bulk quantum effects for de Sitter branes in AdS$_5$}
\end{center} 
\vspace*{4mm} 
\centerline{\large 
Ian G. Moss \footnote{E-mail: ian.moss@ncl.ac.uk}${}^{a}$, 
Wade Naylor \footnote{Current e-mail address: wade@yukawa.kyoto-u.ac.jp}${}^{b}$,
Wenceslao  Santiago-Germ\'{a}n \footnote{E-mail: g.w.santiago-german@ncl.ac.uk}${}^{a}$,  
Misao Sasaki \footnote{Current e-mail address: misao@yukawa.kyoto-u.ac.jp}${}^{b}$}
\vspace*{4mm} 
\centerline{\em ${}^{a}$Department of Mathematics, University of  
Newcastle Upon Tyne,} 
\centerline{\em  Newcastle Upon Tyne, NE1 7RU, UK;} 
\centerline{\em ${}^{b}$Department of Earth and Space Science, Graduate 
School of Science, Osaka University,} 
\centerline{\em Toyonaka 560-0043, Japan} 

\vspace*{4mm}

\begin{abstract}
We investigate some issues regarding quantum corrections for
de Sitter branes in a bulk AdS$_5$ spacetime. The one-loop 
effective action for a Majorana spinor field is evaluated and compared 
with the scalar field result.
We also evaluate the cocycle function for various boundary conditions, 
finding that the quantum corrections naturally induce higher order 
curvature terms in the original action and, in general, it is not possible to 
eliminate the cocycle function by renormalisation. In the one brane 
limit care must be taken on how one extracts physical results. The 
effective potential is found to be zero on the conformally related cylinder. 
However, using the actual metric, the contribution from the cocycle function is
non-zero and must be included. Subtleties with any zero modes are also discussed. 
\vspace*{4mm}

\end{abstract}
\pacs{11.10.Kk, 04.50.+h, 11.25.Mj, 98.80.Cq}

\maketitle
\section{Introduction}

 From within the circumference of brane world cosmology, many ideas 
and proposals relate to the work of Randall and Sundrum \cite{RS1}, who
initially 
implemented their two brane model to try and solve the 
hierarchy problem. In the one brane variant of this model \cite{RS2} it was
also shown 
that the extra dimension could be of infinite extent. In most contexts, 
the second model (and its extensions to 
include curved branes) has been of more interest to the cosmology 
community because gravity behaves essentially like 4-dimensional gravity 
localised near the brane \cite{RS2,SMS,GT}.

An interesting brane world scenario (BWS) has been developed in \cite{KKS,HIME}, 
known as the bulk inflaton model, where it is possible to obtain
inflation on a single positive tension brane solely due
 to the presence of a  bulk gravitational scalar field.
In this regard, the vacuum energy of the bulk scalar field
could have some affect on the cosmological evolution of the
brane, (depending on the size and sign of this quantity). 
The importance of vacuum energy in BWS has also 
been highlighted in \cite{MUKO} where it has been suggested that
it may be possible to solve the hierarchy problem with two positive tension 
branes, if the back-reaction of the Casimir effect is
 included (also see \cite{HKP}).

In this article we evaluate the one-loop effective action for one
and two de Sitter brane configurations in a bulk 5-dimensional
anti-de Sitter spacetime. We focus on Majorana spinor fields and 
compare with the scalar field results. 
This article continues from a previous work \cite{WN} (also see \cite{NOZ} for
the one brane case) for scalar fields. 
As before, we employ $\zeta$-function regularisation \cite{TEN}, 
which is unusual as compared to most regularisation schemes, in that it 
does not require infinite counter terms. Of interest is the recent article by 
Elizalde et.\,al.\,\cite{ENOO}, who also 
considered one and two brane set ups for scalar fields in dS$_5$ as well as
AdS$_5$ 
backgrounds. Small mass perturbations from the massless conformally coupled 
case were also studied. However, here we include the cocycle function (because 
these techniques rely on conformal transformations) which can be related to the 
$B_{5}$ heat kernel coefficient \cite{DA,GPT,GPT2}. 

We briefly mention the work done for a bulk AdS$_5$ background bounded by flat
branes. In \cite{GPT} $\zeta$-function regularisation was employed to evaluate 
the effective potential for a massless conformally coupled scalar and fermion 
field and some special techniques were used for the graviton field. 
The possibility of radion stabilisation due to quantum effects was also
studied, 
which gave a negative result.
The dimensional reduction procedure was used in \cite{TOMS,GR,FT} to obtain an 
effective four dimensional field with an infinite KK tower. The effective
potential was 
then regularised using dimensional regularisation. Fermion fields have also
been 
considered in \cite{FMT,FMT2} where the main emphasis was on fermion 
representations and topological symmetry breaking. Some recent work 
concerning the radion with bulk gauge fields is given in \cite{GP}, \& for 
higher dimensional models, see \cite{FGPT}.

In the next section we discuss the Euclidean de Sitter metric for the bulk
inflaton 
model and the wave equation for scalar and spinor fields.
In section III we make a detailed analysis of the effective potential for a 
conformally coupled Majorana fermion field.
In section IV we evaluate the cocycle function for spin zero fields with
Dirichlet or 
Robin boundary conditions and a spin half field with mixed boundary conditions.
This 
is also evaluated for the delta-function contribution in the potential. 
We discuss the renormalisation of these terms. 
In section V we look at the flat space limit and the Casimir energy. 
In section VI we draw conclusions. In Appendix A we look at the case of twisted 
fields for the Majorana fermion. In Appendix B we discuss zero modes.

\section{de Sitter branes and conformally coupled fields}

We begin with two de Sitter branes embedded in five dimensional anti de Sitter 
space, placed symmetrically to preserve the de Sitter isometry group and 
forming the boundary of the region which we will consider. We will calculate 
the vacuum polarization on a Euclideanised form of the metric. On the 
Euclidean section the anti de Sitter metric becomes a hyperboloid and the de 
Sitter branes become concentric four spheres \cite{GS},
\begin{equation}
ds^2 =dr^2+\ell^2\sinh^2(r/\ell)d\Omega^2_4,
\end{equation}
where $\ell=(-6/\Lambda_5)^{1/2}$ is the anti-de Sitter radius and
$d\Omega_4^2$ is the metric on the unit 4-sphere.
We use $r_-$ for the location of the negative tension brane and $r_+$ for the
location of the positive tension brane ($r_-<r_+$). At the classical level,
boundary or junction conditions relate the locations of the branes to the
brane tensions $\sigma_\pm$ according to \cite{GS,GEN}
\begin{equation}
\sigma_\pm=\pm\frac{3}{4\pi G_5 \ell}\coth(r_\pm/\ell).
\end{equation}
The quantum corrections will introduce other sources of stress-energy on the
brane which modify these relations.

The metric is conformal to a cylinder $I\times S^4$ \cite{GS,GEN}.
However, it may also be of interest to consider the compact space $S^1\times
S^4$ 
without boundaries, in the one brane limit used in the bulk inflaton model
\cite{HIME}, 
and in this case all results are simply multiplied by a factor of two. Thus,
\begin{equation}
ds^2 =a^2(z)(dz^2+d\Omega^2_4)
\quad\qquad a(z)=\frac{\ell}{\sinh(z_0+|z|)}\,,
\label{metric}
\end{equation}
where the coordinates are chosen to have the positive tension brane at $z=0$ 
and the 
negative tension brane at $z=L$, so that the one brane configuration is given
by $L\rightarrow\infty$.
The non-dimensional length $L$ is given by
\begin{equation}
L=\int_{r_-}^{r_+}\frac{dr}
{\ell\sinh(r/\ell)}=
\log\coth{r_-\over 2\ell}-\log\coth{r_+\over 2\ell}.
\end{equation}

The one loop effective action for a conformal scalar field $\phi$ in five 
dimensions on a space with metric $g$ takes the form
\begin{equation}
W[g] = \frac{1}{2}\log\det\Delta[g],
\end{equation}
where $\Delta[g]$ is the conformal operator,
\begin{equation}
\Delta[g]=-\nabla^2+\frac3{16}{}^{5}R
\end{equation}
with Ricci scalar ${}^{5}R$. Conformal symmetry also restricts the boundary 
conditions to be either Dirichlet $\phi=0$ or Robin 
$(-\frac38\theta+\partial_n)\phi=0$, where 
$\theta$ is the trace of the extrinsic curvature.  

If the determinant is defined using zeta-functions, the effective actions on 
conformally related spaces with metrics $g$ and $g_\omega$ are related by
\begin{equation}
W[g_\omega]=C[\Delta,\omega]+W[g],
\label{cocycle}
\end{equation}
where the cocycle function $C[\Delta,\omega]$ can be expressed in terms of
local tensors \cite{DS}. This enables us to reduce the present problem to the
calculation of the cocycle function and the effective action on the cylinder.  

For the cylinder, ${}^{5}R=12$ and
\begin{equation}
\Delta=\left (-\partial_ z{}^2-\Delta^{(4)}+\frac{9}{4}\right),
\label{conformal}
\end{equation}
where $\Delta^{(4)}$ is the Laplacian on the four sphere. The conformal 
symmetry allows either Dirichlet or Neumann boundary conditions and both are 
consistent with a $Z_2$ reflection symmetry about either brane. We shall take
the same boundary conditions on either brane. The spectrum of 
$-\partial_z^2$ in Eq.~(\ref{conformal}) is simply given by $\pi n/L$,  
$n=0,1,2,\cdots$ for Neumann boundary conditions and $n=1,2,\cdots$ for
Dirichlet boundary conditions. (Choosing different boundary conditions on
either brane would give a spectrum $\pi (n+\frac12)/L$, see appendix A.)

We stress that in a more general setting, i.e. $\xi\neq \frac3{16}$, conformal 
transformations introduce delta functions in the potential, Eq.
(\ref{conformal}), 
because the scale factor $a(z)$, Eq. (\ref{metric}), depends on the absolute
value $|z|$. 
Only conformal coupling to the curvature, $\xi=\frac3{16}$, irrespective of any
mass term 
cancels out any distributional sources. This is for the same reason that
Neumann 
boundary 
conditions remain Neumann after performing a conformal transformation of the 
field. In terms of the bulk inflaton model \cite{HIME} for $\xi=\frac3{16}$
there is no 
bound state and gravity cannot be localised on the brane. However, the cocycle 
function interpolates between the original metric and the scaled one and
therefore 
introduces distributional terms. 

The eigenvalues of the Laplacian $\Delta^{(4)}$ are $m(m+2)$ with degeneracy
\begin{equation}
d(m)=\frac{1}{3}(m+2)(m+3/2)(m+1)=
\frac{1}{3}\left[(m+3/2)^3-\frac{1}{4}(m+3/2)\right],
\label{deg}
\end{equation}
where we have rearranged this expression for later convenience.
Thus, the eigenvalues for the Klein-Gordon operator, Eq. (\ref{conformal}), are
\begin{eqnarray}
\lambda_{n,m}=\left(\frac{\pi n}{L}\right)^2+(m+3/2)^2\,.
\end{eqnarray}

For spin-$1/2$ fermion fields $\psi$, the massless Dirac equation is 
automatically conformally covariant. The boundary conditions force exactly 
half of the fermion components to vanish at the brane, and generally take the 
form
\begin{equation}
\frac12(1+S)\psi=0,
\end{equation}
where $S$ is a spinor transformation representing reflection about the 
boundary. We shall concentrate on Majorana spinors, which have 8 real
components. Reflections about the boundaries can be represented by using
$8\times 8$ $\Gamma$ matrices, $S=i\Gamma_5\Gamma_6$ . The details can be
found in reference \cite{FMT}.

It is convenient to take the square of the Dirac operator when 
evaluating the effective action
\begin{equation}W=-\frac{1}{4}\log\det\Delta\end{equation}
for Majorana fermions, where
\begin{equation}
\Delta=-\nabla^2+\frac14{}^{5}R.
\end{equation}
The squared operator is not conformally invariant, but it is nevertheless 
still possible to relate the effective actions of fermions with conformally 
related metrics \cite{MOSS2}.
 
On the cylinder,
\begin{equation}
\Delta=\left (-\partial_ z{}^2-\Delta^{(4)}_f+3\right).
\label{eigferm}
\end{equation}
The eigenvalues of the spinor Laplacian, $\Delta^{(4)}_f$, on the 4-sphere 
are well known (e.g., see \cite{MOSS}) and are given by $(m+2)^2-3$. 
Half of the field components satisfy Dirichlet boundary conditions and half 
satisfy Neumann boundary conditions, and the eigenvalues of $\Delta$ are
\begin{equation}
{\lambda_{n,m}^M}=\left(\frac{\pi n}{L}\right)^2+(m+2)^2\,.
\end{equation}
The degeneracy for the 8 component Majorana spinors is given by
\cite{MOSS}
\begin{equation}
d^M(m)=8\times\frac{1}{6}(m+1)(m+2)(m+3)=\frac{4}{3}\left[(m+2)^3-(m+2)\right],
\label{degferm}
\end{equation}
where we take $n\in Z$.

\section{Conformally flat scalar and spinor fields}

\subsection{The effective action and $\zeta$ functions}

We can employ the $\zeta$-function method to find the contribution to the 
effective action from the cylinder. The generalised zeta function is given by
\begin{equation}
\label{gfunc}
\zeta(s)=\sum_{m,n=0}^\infty d(m) \lambda_{m,n}^{-s}
\end{equation}
with the one-loop effective action related to $\zeta(s)$ by (e.g., see 
\cite{BD,MOSS})
\begin{equation}
W=-\frac{1}{2}\,\zeta'(0)-\frac{1}{2} \zeta(0) \log \mu^2
\label{effscal}
\end{equation}
for scalar fields and
\begin{equation}
\label{effact}
W=\frac{1}{4}\,\zeta'(0)+\frac{1}{4} \zeta(0) \log \mu^2
\end{equation}
for Majorana fermions, where $\mu$ is the renormalisation scale.

The scalar field result is contained in previous work \cite{WN} and in the 
early work of Dowker and Apps \cite{DA}. In what follows, we therefore 
concentrate on the spinor field. The scalar field results are quoted for
comparison (see appendix B).

\subsection{One brane}
We first evaluate the effective potential for a one brane configuration on the
cylinder, where $L\rightarrow\infty$ and the discrete sum over $n$ becomes an
integral. It is then simple to see that the zeta function $\zeta^M(s)$ becomes
\begin{equation}
\zeta^M(s)=\frac{2L}{\pi}\int_0^\infty dk \sum_{m=0}^\infty d^M(m)
\left(k^2+(m+2)^2\right)^{-s},
\end{equation}
where the factor of $2$ is because there are two copies of the bulk space on
either side of the brane, essentially we assume an $S^1\times S^4$ topology. 
For large $s$ we can interchange the order of the sum and the integral and 
perform the $k$ integration, 
using the well known identity
\begin{equation}
\int_0^\infty dk k^\alpha (k^2+a)^{-s}=\frac{1}{2}
\Gamma\left(\frac{\alpha+1}{2}\right)
\frac{\Gamma(s-1/2-\alpha/2)}{\Gamma(s)}\,a^{\alpha/2+1/2-s},
\end{equation}
implying
\begin{eqnarray}
\zeta^M(s)&=&\frac{2L}{\pi}\,
\frac{\sqrt{\pi}}{2}\sum_{m=0}^\infty
\frac{\Gamma(s-1/2)}{\Gamma(s)}\;d^M(m)(m+2)^{1-2s},\nonumber\\
&=&\frac{L}{\pi}\frac 4 3\sqrt{\pi}\frac{\Gamma(s-1/2)}{\Gamma(s)}
\left(\zeta(2s-4,2)-\zeta(2s-2,2)\right),
\end{eqnarray}
where in the second step we have used simple algebra to rewrite the equation in
terms 
of generalised (Hurwitz) $\zeta$-functions. For $s=0$ it is clear that
$\zeta^M(0)=0$ 
because
\begin{equation}
\label{gam}
\frac{1}{\Gamma(s)}=s+\gamma s^2+O(s^3),
\end{equation}
where $\gamma$ is Euler's constant. Thus,
\begin{equation}
\label{1bravac}
\zeta^{M\prime}(0)=\frac{L}{\pi}\frac 4 3\sqrt{\pi}\,\Gamma(-1/2)
\left(\zeta(-4,2)-\zeta(-2,2)\right)=0,
\end{equation}
which is zero because the combination of $\zeta$-functions
in Eq. (\ref{1bravac}) cancel, as can be
verified by employing the relation
\begin{equation}
\zeta(-n,a)=-\frac{B_{n+1}(a)}{n+1}\,,
\label{bern}
\end{equation}
where $B_n(x)$ is a Bernoulli polynomial.
Therefore, as for scalar fields \cite{WN,NOZ}, the one-loop effective potential
on the 
cylinder is zero in the one brane case. We discuss the cocycle contribution
later.

\subsection{Two branes}
We now consider the two brane configuration which consists of
two infinite summations, which can be expressed in
terms of generalised $\zeta$-functions. In fact,
there are many subtle issues regarding the correct analytic
continuation of such functions upon interchanging the order of the summations
(see \cite{TEN} for a detailed discussion).
As we shall see, the one brane result, above, appears as one of the terms
in the two brane case. Here assume the topology is $S^1\times S^4$ rather than 
$I\times S^4$, to compare with the result of the last section. 
The function $\zeta^M(s)$ now becomes
\begin{eqnarray}
\zeta^M(s)&=& \sum_{n=-\infty}^\infty\,
\sum_{m=0}^\infty d^M(m)\left(c^2 n^2+(m+2)^2\right)^{-s},
\nonumber\\
\label{2bravac}
&=&\frac{1}{\Gamma(s)} \sum_{n=-\infty}^\infty\,
\sum_{m=0}^\infty d^M(m)\int_0^\infty
dt\;t^{s-1} \exp\{-t[\;c^2n^2+(m+2)^2]\},
\end{eqnarray}
where $c=\pi/L$ and in the second step we have made use
of the Mellin transform. The above case refers to an untwisted 
spinor field, the twisted case is discussed in the appendix.
This formula is almost identical in form to the
2-dimensional Epstein-Hurwitz $\zeta$-function
studied by Elizalde in \cite{ELI2} (see also \cite{TEN}),
apart from the spinor degeneracy factor $d^M(m)$.
This work suggests using the transformation formula of
Jacobi's $\theta_3$ function \cite{BATE}, equivalent to a Poisson resummation,
\begin{equation}
\sum_{n=-\infty}^\infty\;e^{-c^2n^2}=\theta_3(0,e^{-c^2})
=\sqrt{\frac{\pi}{c^2}}\,\theta_3(0,e^{-\pi^2/c^2})=\sqrt{\frac{\pi}{c^2}}
\sum_{n=-\infty}^\infty\,\exp[-\pi^2n^2/c^2].
\label{poiss}
\end{equation}

Substitution of the above equation into (\ref{2bravac})

allows us to interchange the order of the summations, giving
\begin{eqnarray}
\zeta^M(s)&=&\sqrt{\frac{\pi}{c^2}}\frac{1}{\Gamma(s)}
\sum_{m=0}^\infty  d^M(m)\int_0^\infty dt\,t^{s-3/2}\exp\{-t(m+2)^2\}
\nonumber\\
&&+2\sqrt{\frac{\pi}{c^2}}\frac{1}{\Gamma(s)}\sum_{n=1,\;m=0}^\infty
\;d^M(m)\int_0^\infty dt\,t^{s-3/2} \exp\left[-\frac{\pi^2 
n^2}{c^2t}-(m+2)^2t\right].
\label{halfway}
\end{eqnarray}
After integrating with respect to $t$ (using standard relations) we obtain
\begin{eqnarray}
\zeta^M(s)&=&\frac{4}{3}\sqrt{\frac{\pi}{c^2}}\frac{\;\Gamma(s-1/2)}{\Gamma(s)}
\left(\zeta(2s-4,2)-\zeta(2s-2,2)\right)
\nonumber\\
&&+\frac{4\pi^s}{\Gamma(s)}\,c^{-s-1/2}\sum_{n=1,\;m=0}^\infty
\;d^M(m)\;n^{s-1/2}(m+2)^{-s+1/2}\,
K_{s-1/2}[2\pi c^{-1}n(m+2)],
\end{eqnarray}
where $K_\nu$ is a modified Bessel function of the second kind.
It is easy to verify that $\zeta^M(0)=0$, using Eq. (\ref{gam}).
Furthermore, taking the derivative of $\zeta^M(s)$ with respect to $s$
(and leaving only terms that
remain independent of $\Gamma(s)$ near $s=0$, see Eq. (\ref{gam})) gives
\begin{eqnarray}
\label{epst}
\zeta^{M\prime}(0)&=&\frac{4}{3}\sqrt{\frac{\pi}{c^2}}
\Gamma(-1/2)\left(\zeta(-4,2)-\zeta(-2,2)\right)
\nonumber\\
&&+4c^{-1/2}\sum_{n=1,\;m=0}^\infty
\;d^M(m)\;n^{-1/2}(m+2)^{1/2}\;K_{-1/2}[2\pi c^{-1}n(m+2)],
\end{eqnarray}
where a prime denotes differentiation with respect to $s$.
Interestingly, the first term is exactly the contribution from the
one brane configuration
Eq. (\ref{1bravac}), which is zero. As mentioned in \cite{WN} this is similar
to finite 
temperature field theory where the $n=0$ mode gives the zero temperature
contribution. 
The similarity to finite temperatures is due to the compact topology in the
fifth dimension.
 
The second term is non-zero and depends on the value of $c=\pi/L$.
Simplifying the above equation we find
\begin{equation}
\frac{1}{4}\,\zeta^{M\prime}(0)=\sum_{n=1,\;m=0}^\infty
\,\frac{2}{3}\frac{(m+1)(m+2)(m+3)}{n}
\exp[-2Ln(m+2)],
\end{equation}
where we have used the relation $K_{-1/2}(z)=\sqrt{\pi/(2z)}\,e^{-z}$. After
evaluating the summation over $m$ we have an expression for the Majorana
spinor one-loop effective action on the cylinder (for $I\times S^4$) 
\begin{equation}
W^{M}_{I\times S_4}=\frac{1}{8}
\sum^\infty_{n=1}\frac{1}{n(\sinh nL)^4},
\label{afl}
\end{equation}
where we have divided by two, since $S^1\times S^4$ doubles the modes.

For the scalar field we only need to make simple replacements in the degeneracy
and 
eigenvalues, see \cite{WN}.
Dowker and Apps gave the first results for scalar fields on a cylinder
\cite{DA}, including the zero mode (see appendix B). Their results extend
trivially 
to five dimensions, with
\begin{equation}
W^{D,N}_{I\times S_4}=\pm\frac14\zeta'_{S_4}(0)
\pm\frac14\zeta_{S_4}(0)\log\mu^2
-\frac{1}{16}\sum^\infty_{n=1}\frac{\cosh nL}{n(\sinh nL)^4}
\end{equation}
for Dirichlet (D) and Neumann (N) boundary conditions, where $\zeta_{S_4}$ is
the zeta function for the operator $-\Delta^{(4)}+\frac{3}{16}\,^5R$ on $S_4$.
Explicitly,
\begin{equation}
\zeta_{S_4}(0)=-\frac{17}{2880},\qquad
\zeta'_{S_4}(0)=\left(-\frac7{24}\zeta_R'(-3)+\frac1{24}\zeta_R'(-1)
+\frac1{24}\zeta_R(-3)\log2-\frac1{24}\zeta_R(-1)\log2\right)\label{zpsphere},
\end{equation}
where $\zeta_R$ is the Riemann zeta function (see appendix B).

\section{The cocycle function}

Calculating the cocycle function is heavily dependent on knowing appropriate 
heat kernel coefficients. These heat kernel coefficients are defined in $d$ 
dimensions by an asymptotic expansion,
\begin{equation}
{\rm tr}\left(\omega e^{-\Delta t}\right)\sim t^{-d/2}\sum_{n=0}^\infty
B_n[\Delta,\omega]t^{n/2}.
\end{equation}
In five dimensions we require $B_5[\Delta,\omega]$, which can be found in the 
literature \cite{GILK}. It takes the form of a surface integral of scalar 
invariants with a maximum of four derivatives of the metric.

Given a sequence of metrics $g_\epsilon=e^{-2\epsilon\omega}g$, and operators
$\Delta_\epsilon$, it can be shown that \cite{DS}
\begin{equation}   
W[g_1]-W[g]=C[\Delta,\omega]=\int_0^1 d\epsilon 
B_d[\Delta_\epsilon,\omega].
\end{equation}
To keep things simple, we have restricted attention to the class of metrics of 
the form
\begin{equation}
g_\epsilon=e^{-2\epsilon\omega(z)}\left(dz^2+d\Omega_4^2\right)
\end{equation}
with boundaries at $z=0$ and $z=L$. The cocycle function then takes a generic 
form
\begin{equation}
C[\Delta,\omega]={1\over 16\pi^2}\int_{S^4}\left\{
\alpha_0\omega+\alpha_1\omega_{zzzz}+\alpha_2\omega_{zzz}\omega_{z}+
\alpha_3\omega_{zz}^2+\alpha_4\omega_{z}^2\omega_{zz}+\alpha_5\omega_{z}^4+
\alpha_6\omega_{z}^2+\alpha_7\omega_{zz}\right\}.
\label{fullc}
\end{equation}
The coefficients are tabulated in table \ref{taba1}. In particular, notice 
that we have evaluated the contribution to the cocycle function for the 
delta function background. This is simply obtained by adding the 
Dirichlet 
contribution to the Robin contribution, see \cite{GKV}. 

Anti de Sitter space has
\begin{equation}
\omega=\log{\sinh(|z|+z_0)\over \ell}=-\log\left(\ell\sinh(r/\ell)\right),
\end{equation}
where we ignore absolute value of $z$, $|z|$, because we have already 
included its effect (delta function terms) in the cocycle function. 
Using the relation $z=-\log \tanh(r/2\ell)$ to transform back to the 
spherical coordinate system, the cocycle function is then 
\begin{equation}
C[\Delta,\omega]=\frac16\sum_{\pm}\left\{
c_0+c_2\sinh^2(r_\pm/\ell)+c_4\sinh^4(r_\pm/\ell)-\alpha_0\log
\left(\ell\sinh(r_\pm/\ell)\right)\right\}.
\end{equation}
The coefficients are tabulated in table \ref{taba2}. Results for scalar fields
with Dirichlet boundary conditions have also been obtained by Garriga et. al.
\cite{GPT2}.

\begin{table}
\caption{Coefficients of the terms in the cocycle function for conformal fields
on Anti de Sitter space with spin 0 and 1/2 with Dirichlet (D),  Robin (R) and
mixed (M) boundary conditions. The spinors are Majorana. 
$\delta$ is the contribution due to the delta-function potential instead of a
boundary.}
\renewcommand{\arraystretch}{1.5}
\begin{ruledtabular}
\begin{tabular}{rrrrrrrrrr}
spin&background&$\alpha_0$&$\alpha_1$&$\alpha_2$&$\alpha_3$&$\alpha_4$
&$\alpha_5$&$\alpha_6$&$\alpha_7$
\\\hline
$0$&D&$\frac{17}{1920}$&$-\frac1{128}$&$\frac5{192}$
&$\frac1{48}$&$-\frac{9}{768}$&$-\frac{11}{3072}$&$-\frac1{16}$&$\frac1{64}$\\
$0$&R&$-\frac{17}{1920}$&$\frac1{128}$&$-\frac1{48}$
&$-\frac1{48}$&$\frac{5}{768}$&$\frac1{15360}$&$\frac1{32}$&$-\frac1{64}$\\
$1/2$&M&$0$&$0$&$\frac1{48}$&$0$&$-\frac1{96}$&$-\frac{11}{1280}$
&$-\frac3{32}$&$0$\\
$0$&$\delta$&$0$&$0$&$\frac1{192}$&$0$&$-\frac1{192}$
&$-\frac{9}{2560}$&$-\frac1{32}$&$0$\\
\end{tabular}
\end{ruledtabular}
\label{taba1}
\end{table}

\begin{table}
\caption{Coefficients of the terms in the cocycle function for conformal fields
on Anti de Sitter space with spin 0 and 1/2 with Dirichlet (D),  Robin (R) and
mixed (M) boundary conditions. The spinors are Majorana. 
$\delta$ is the contribution due to the delta-function potential instead of a
boundary.}
\renewcommand{\arraystretch}{1.5}
\begin{ruledtabular}
\begin{tabular}{rrrrr}
spin&background&$c_0$&$c_2$&$c_4$\\
\hline
$0$&D&$-\frac{203}{3072}$&$\frac{30}{3072}$&$\frac{393}{3072}$\\
$0$&R&$\frac{481}{15360}$&$-\frac{498}{15360}$&$-\frac{1779}{15360}$\\
$1/2$&M&$-\frac{393}{3840}$&$-\frac{226}{3840}$&$\frac{167}{3840}$\\
$0$&$\delta$&$-\frac{89}{2560}$&$-\frac{58}{2560}$&$\frac{31}{2560}$\\
\end{tabular}
\end{ruledtabular}
\label{taba2}
\end{table}

It is important to discuss whether any of the terms in the cocycle function can
be regarded as renormalisations of the original action. Consider first of all
the terms with coefficient $c_4$. These terms are proportional to the volumes
of the branes since, for example, the positive tension brane is a four-sphere
of radius $\ell\sinh(r_+/\ell)$. We have only considered bulk fields, but in a
more realistic theory there would also be matter fields on the brane which 
could give infinite renormalisations of the brane tensions. It would then make
sense to absorb $c_4$ into a finite renormalisation. However, when there are
two branes, the new terms take the same sign on the positive tension and the
negative tension brane and it would be very unnatural to then suppose that the
renormalised brane tensions could be exactly equal and opposite sign as they
are in the Randall-Sundrum model. 

The terms with coefficients $c_2$ and $c_0$ are similar to brane curvature
terms ${}^4R$ and curvature squared terms in the Lagrangian. Such terms may be
desirable to renormalise away divergences from the brane fields, but they
represent a significant departure from the models which we mentioned in the
introduction. Both the classical field equations and the quantisation would be
modified. A possible resolution would be to regard the theory as a low energy
approximation and restrict attention to theories in which the divergences from
the brane fields vanish at one loop order.

Another feature becomes apparent when we examine the coefficients in equation
(\ref{fullc}), which imply that the cocycle function combines brane curvature
with extrinsic curvature terms. These can only be expressed in terms of brane
curvature terms (by the Gauss-Codacci relations) in special cases. We conclude
that the cocycle function should not be eliminated by renormalisation.

Combining the cocycle function with the cylinder result gives the total one
loop effective action for the Majorana fermion,
\begin{equation}
W^M=\frac16\sum_{\pm}\left\{
c^M_0+c^M_2\sinh^2(r_\pm/\ell)+c^M_4\sinh^4(r_\pm/\ell)\right\}
+\frac{1}{8}\sum^\infty_{n=1}\frac{1}{n(\sinh nL)^4}.
\label{wm}
\end{equation}
where the coefficients are tabulated in table \ref{taba2}. For comparison, the
conformal scalar effective 
actions are
\begin{eqnarray}
W^{D,R}&=&\frac16\sum_{\pm}\left\{
c^{D,R}_0+c^{D,R}_2\sinh^2(r_\pm/\ell)+c^{D,R}_4\sinh^4(r_\pm/\ell)
-\alpha^{D,R}_0\log\left(\ell\mu\sinh(r_\pm/\ell)\right)\right\}\nonumber\\
&&\pm\frac14\zeta'_{S_4}(0)-\frac{1}{16}\sum^\infty_{n=1}\frac{\cosh nL}{n(\sinh
nL)^4}.
\label{ws}
\end{eqnarray}
for Dirichlet and Robin boundary conditions, where $\zeta'_{S_4}(0)$ was given
in equation (\ref{zpsphere}).

\section{Flat space limit and Casimir energy density}

The results can be simplified in a variety of limiting cases. The curvatures of 
the branes are small when $r_\pm\gg\ell$ and they approximate the flat branes 
of the original Randall-Sundrum model. The conformal distance $L$ becomes 
small,
\begin{equation}
L\sim 2\left(e^{-r_-/\ell}-e^{-r_+/\ell}\right)
\end{equation}

The effective action for spin 1/2 fields, including the cocycle function, is 
then
\begin{equation}
W\sim {1\over 96}c_4\left(e^{4r_+/\ell}+e^{4r_-/\ell}\right)- {\zeta_R(5)\over
16}L^{-4}.
\end{equation}
In effect, there is an effective potential on the negative tension brane
defined by
\begin{equation}
V={3 W\over 8\pi^2\ell^4\sinh^4(r_-/\ell)}
\end{equation}
We can express this potential in terms of a radion field 
$\sigma=(r_+-r_-)/\ell$,
\begin{equation}
V\sim{1\over 16 \ell^4}c_4\left(1+e^{4\sigma}\right)
-A\left(e^\sigma-1\right)^{-4}
\end{equation}
where
\begin{equation}
A=-{3\zeta_R(5)\over 128 \ell^4}
\end{equation}
The result agrees with the flat space Casimir energy for spin 1/2 fields 
calculated previously \cite{GPT,FMT,FMT2}. Some of these old results absorbed
the
first two terms into shifts in the brane tension. 

Another limiting case is when the brane separation is small,  
$(r_+-r_-)\ll \ell\sinh(r_-/\ell)$. The conformal separation becomes
\begin{equation}
L\sim{r_+-r_-\over \ell\sinh(r_-/\ell)}
\end{equation}
and is small in this limit. The Casimir energy
\begin{equation}
V\sim-A\sigma^{-4}
\end{equation}
is independent of the five dimensional cosmological constant.

Finally, the limit $r_-\ll \ell$ corresponds to large conformal separation and
approaches the single brane limit. In this case
\begin{equation}
L\sim -\log\left({r_-\over 2\ell}\right)-\log\coth{r_+\over 2\ell}
\end{equation}
The contribution to the effective action from the cylinder disappears as
$r_-^{\,4}$ and we are left only with the cocycle function on the positive
tension brane. The effective potential is then obtained by dividing by the
finite 
5-dimensional volume (using the actual metric) and this contribution still 
remains. Note, on the cylinder the effective potential is zero, in the 
one brane limit, because the conformal volume is infinite, i.e., the 5D volume 
is proportional to $1/L$.

\section{Conclusion}

We have discussed the effective action on curved brane backgrounds for
conformally coupled scalar and spinor fields. The final formulae, which can be
found at the end of section 4, consist of a contribution from a region of the
cylinder and a cocycle function resulting from a conformal rescaling of the
background.

Perturbations around the conformal case from a mass have been considered in
\cite{ENOO} on the cylinder background. These terms make small corrections to
the effective action. A mass, or small correction to the conformal coupling,
will also contribute to the cocycle function. However, its generic form does
not change from the case presented here. 

As we have seen, the cocycle function induces curvature terms and, in general,
it is not possible to eliminate the cocycle function by renormalisation.
However, some parts can be absorbed by renormalisation, for example as finite
renormalisation of the brane tensions. Other parts of the cocycle function are
additional curvature terms. It has been known for some time that quantum
fields on curved brane backgrounds give induced curvature terms (for e.g., see
\cite{FT} for induced terms in the RS model).

On the conformally related cylinder, the effective potential vanishes in the
single brane limit. This is expected because, as first argued in \cite{NOZ},
the background becomes a half-cylinder with infinite volume.
However, using the actual metric the effective potential includes a
contribution from the cocycle function, as we have shown. 
Note that the volume of the spacetime (after the usual Wick rotation of the
time) is finite. Therefore, we must be careful in interpreting any result on
the conformally related cylinder. 
For a two brane configuration, the effective action on the cylinder is
non-zero, because the cylindrical region is compact, and the cocycle function
does not vanish. 

In tables \ref{taba1} \& \ref{taba2} we also separated out the part due to the 
delta function in the background potential. A puzzling point is that the result
is not the same as what one would obtain for a thin but finite domain wall,
for which the cocycle function would vanish. Hence the thin wall limit
will not agree with the case of the delta function potential.
The physical reason behind this discrepancy is unclear.
We leave this issue for future study.
Until it is resolved, the result for the delta function potential
should perhaps be taken with care.
In this connection, we note that the thin wall limit in bubble nucleation
was analyzed in \cite{MN}, and no anomalous behaviour was found.

When considering backreaction for conformally coupled fields, the
semi-classical Einstein equations are exact and the Casimir energy can play an
important role in BWS \cite{MUKO,HKP}, including inflating branes. 
For flat branes this only depends on the difference between the number of
bosons ($N_b$) and fermions ($N_f$), where $N_b=N_f$ gives the usual RS model
and $N_b>N_f$ gives two positive tension branes, see \cite{MUKO}.
For de Sitter branes we clearly see (compare Eq. (\ref{wm}) with 
Eq. (\ref{ws})\,) that in general the contribution from scalar and 
fermion fields are not the same, as was argued in \cite{WN}. 
Thus, in this case, the bulk will not remain pure anti-de Sitter under
back-reaction even for $N_b=N_f$.

As well as untwisted fields we also considered twisted fields (see appendix
A). This case occurs when one brane obeys Dirichlet and the other Neumann
boundary 
conditions. In fact,  {\it a priori}, there seems to be no requirement that a
scalar field 
should only satisfy untwisted boundary conditions. One could then envisage a
two brane 
set up with the positive tension brane satisfying Neumann and the other
negative 
tension brane satisfying Dirichlet boundary conditions. If this were the case
then 
it should be possible to have a bound state localised on the positive tension
brane even 
in a two brane set up.

\acknowledgements{
Many thanks to James Norman for checking the cocycle function
results. W.N. and M.S. wish to thank Antonino Flachi and Yoshiaki Himemoto for 
enlightening discussions.
W.N. acknowledges support from JSPS for Postdoctoral Fellowship for Foreign 
Researchers No. P01773.
W.S. is grateful to CONACYT of Mexico, Grant Number 116020, for financial 
support.
The work of M.S. is supported by Monbukagaku-sho Grant-in-Aid
for Scientific Research (S) No. 14102004.}

\appendix

\section{Twisted fields}
For a spinor field it is also possible to have twisted as well as untwisted 
field configurations (see \cite{FMT} for the flat brane case). In Section IIIC
we considered the untwisted case. 
Here, we will double the modes, i.e. working on the topology $S^1\times S^4$. 
In any case the zero modes cancel for mixed boundary conditions (see appendix
B).
The function $\zeta^{TM}(s)$, Eq. (\ref{2bravac}), now becomes (with 
mixed boundary conditions)
\begin{eqnarray}
\zeta^{TM}(s)&=& \sum_{n=-\infty}^\infty\,
\sum_{m=0}^\infty d(m)\left(c^2 (n+1/2)^2+(m+2)^2\right)^{-s},
\nonumber\\
\label{2bratwst}
&=&\frac{1}{\Gamma(s)} \sum_{n=-\infty}^\infty\,
\sum_{m=0}^\infty d^M(m)\int_0^\infty
dt\;t^{s-1} \exp\{-t[\;c^2(n+1/2)^2+(m+2)^2]\}.
\end{eqnarray}
We can express the twisted result in terms of 
the known result for the untwisted case by using the identity \cite{BATE}
\begin{equation}
\theta_2(0,e^{-t})=\theta_3(0,e^{-t/4})-\theta_3(0,e^{-t}),
\end{equation}
where $\theta_3(0,e^{-t})$ is defined in Eq. (\ref{sum}). (This procedure is 
very similar to that used in finite temperature field theory, when considering
thermal 
bosons and fermions, e.g., see \cite{ALLEN}.) Thus, applying the same steps as
we did for the 
untwisted case (see the steps following Eq. (\ref{2bravac}))  we obtain
\begin{eqnarray}
\zeta^{TM\prime}(0)&=&
\frac{4}{3}\sqrt{\frac{\pi}{c^2}}
\Gamma(-1/2)\left(\zeta(-4,2)-\zeta(-2,2)\right)
\nonumber\\
&&-4c^{-1/2}\sum_{n=1,\;m=0}^\infty
\;d^M(m)\;n^{-1/2}(m+2)^{1/2}\;K_{-1/2}[2\pi c^{-1}n(m+2)],
\nonumber\\
&&+4\sqrt{2}c^{-1/2}\sum_{n=1,\;m=0}^\infty
\;d^M(m)\;n^{-1/2}(m+2)^{1/2}\;K_{-1/2}[4\pi c^{-1}n(m+2)].
\end{eqnarray}
The one brane contribution is zero and the effective action is given by
\begin{equation}
W^{TM}_{I\times S_4}=\frac{1}{8}\sum^\infty_{n=1}\frac{1}{n(\sinh 2nL)^4}-
\frac{1}{8}\sum^\infty_{n=1}\frac{1}{n(\sinh nL)^4},
\end{equation}
where we have divided by two to obtain the result on $I\times S^4$.
It is simple to show that this result agrees with 
the flat brane result given in \cite{FMT} for a twisted spinor field, by 
simply taking the small $L$ limit.

\section{Zero modes}
Here we discuss a slight subtlety with boundary conditions. Note, by zero modes 
we mean the $n=0$ modes in our mode sum and not null eigenvectors.
In \cite{WN} we considered the case of a scalar field, but 
neglected the zero modes because from the one brane 
point of view it is more convenient to consider $S^1\times S^4$, which has no 
boundaries. 
For the mode sum, over $n$, we argued that we should double the modes given 
that there are two copies of the bulk on either side of the brane, 
required to obtain a compact spacetime. However, from Elizalde \cite{ELI2},
\begin{equation}
\sum_{n=0}^\infty\;e^{-c^2n^2}=\mp\frac{1}{2}+\frac{1}{2}\sqrt{\frac{\pi}{c^2}}
+\sqrt{\frac{\pi}{c^2}}\sum_{n=1}^\infty\,\exp[-\pi^2n^2/c^2],
\end{equation}
where the $\mp$ refers to Dirichlet and Neumann boundary conditions
respectively. 
For the case of a spinor field with mixed boundary conditions (with topology 
$I\times S^4$) these modes cancel each other out. Thus, 
doubling the modes in the above equation we obtain
\begin{equation}
\sum_{n=-\infty}^\infty\;e^{-c^2n^2}=\mp\,1+
\sqrt{\frac{\pi}{c^2}}\sum_{n=-\infty}^\infty\,\exp[-\pi^2n^2/c^2].
\label{sum}
\end{equation}
This corresponds to $2I\times S^4$. In the case of $S^1\times S^4$ the above 
equation reduces to Eq. (\ref{poiss}), i.e. no zero modes.

As before, using Eq. (\ref{2bravac}),
\begin{eqnarray}
\zeta(s)&=&\mp\sum_{m=0}^\infty d(m)(m+3/2)^{-2s}
+\sqrt{\frac{\pi}{c^2}}\frac{1}{\Gamma(s)}
\sum_{m=0}^\infty  d(m)\int_0^\infty dt\,t^{s-3/2}\exp\{-t(m+3/2)^2\}
\nonumber\\
&&+2\sqrt{\frac{\pi}{c^2}}\frac{1}{\Gamma(s)}\sum_{n=1,\;m=0}^\infty
\;d(m)\int_0^\infty dt\,t^{s-3/2} \exp\left[-\frac{\pi^2 
n^2}{c^2t}-(m+3/2)^2t\right],
\end{eqnarray}
where in the first term, we do not need to make use of the Mellin transform. 
Then integrating with respect to $t$ 
\begin{eqnarray}
\zeta(s)&=&\mp \zeta_{S_4}(s)
+\sqrt{\frac{\pi}{c^2}}\frac{\;\Gamma(s-1/2)}{\Gamma(s)}\zeta_{S_4}(s-1/2)
\nonumber\\
&&+\frac{4\pi^s}{\Gamma(s)}\,c^{-s-1/2}\sum_{n=1,\;m=0}^\infty
\;d(m)\;n^{s-1/2}(m+3/2)^{-s+1/2}\,
K_{s-1/2}[2\pi c^{-1}n(m+3/2)],
\end{eqnarray}
Again, apart from $\zeta_{S_4}(s)$, it is easy to verify that $\zeta(0)=0$.
Whence,
\begin{eqnarray}
\zeta^{\prime}(0)&=&\mp\zeta_{S_4}{}'(0)
+\frac{1}{3}\sqrt{\frac{\pi}{c^2}}
\Gamma(-1/2)\left(\zeta(-4,3/2)-\zeta(-2,3/2)\right)
\nonumber\\
&&+4c^{-1/2}\sum_{n=1,\;m=0}^\infty
\;d(m)\;n^{-1/2}(m+2)^{1/2}\;K_{-1/2}[2\pi c^{-1}n(m+3/2)],
\end{eqnarray}
where a prime denotes differentiation with respect to $s$.
As expected the second term is exactly the contribution from the
one brane configuration, where we have used the identity,
\begin{equation}
\frac{d}{ds}\left[
\frac{2}{\sqrt{4\pi}}\frac{\;\Gamma(s-1/2)}{\Gamma(s)}\zeta_{S_4}(s-1/2)\right]_{s=0}
=\frac{d}{ds}\left[\frac{1}{3}\frac{L}{\sqrt{\pi}}\frac{\;\Gamma(s-1/2)}{\Gamma(s)}
\left(\zeta(2s-4,3/2)-\frac{1}{4}\zeta(2s-2,3/2)\right)\right]_{s=0}=0,
\end{equation}
with $c=\pi/L$. Thus, in the one brane limit the effective potential reduces to
the Casimir 
energy on $S^4$, as pointed out in \cite{NOZ}. This relation can also be found
in 
\cite{DA}, except for a factor of $2$ because of our mode doubling.

The third term is non-zero and depends on the value of $c=\pi/L$,
\begin{equation}
\frac{1}{2}\,\zeta^{\prime}(0)=\sum_{n=1,\;m=0}^\infty
\,\frac{1}{3}\frac{(m+1)(m+2)(m+3/2)}{n}
\exp[-2Ln(m+3/2)],
\end{equation}
where we have used the relation $K_{-1/2}(z)=\sqrt{\pi/(2z)}\,e^{-z}$ and
Eq. (\ref{deg}). After evaluating the summation over $m$ we have an expression 
for the scalar one-loop effective action on the cylinder,
\begin{equation}
W_{2I\times S_4}=\frac{1}{8}
\sum^\infty_{n=1}\frac{\cosh nL}{n(\sinh nL)^4}.
\end{equation}
Thus, since the one brane contribution is zero,
\begin{equation}
W^{D,N}_{2I\times S_4}=\pm\frac12\zeta'_{S_4}(0)
\pm\frac12\zeta_{S_4}(0)\log\mu^2
-\frac{1}{8}\sum^\infty_{n=1}\frac{\cosh nL}{n(\sinh nL)^4}=2\,W^{D,N}_{I\times S_4},
\end{equation}
with Dirichlet (D) and Neumann (N) boundary conditions, with
\begin{equation}
\zeta_{S_4}(s)=\frac{1}{3}\left(\zeta(2s-3,3/2)
-\frac{1}{4}\,\zeta(2s-1,3/2)\right),
\label{s4zeta}
\end{equation}
where $\zeta(s,a)$ is a Hurwitz $\zeta$-function. Thus,
\begin{equation}
\zeta_{S_4}(0)=\frac{1}{3}\left(\zeta(-3,3/2)
-\frac{1}{4}\,\zeta(-1,3/2)\right)
=-\frac{17}{2880}
\end{equation}
and using the standard relation
\begin{equation}
\zeta'(s,3/2)=2^s\log 2\zeta(s)+(2^s-1)\zeta'(s)-2^s\log2
\end{equation}
we find that
\begin{eqnarray}
\zeta'_{S_4}(0)&=&\frac{1}{3}\left(\zeta'(-3,3/2)
-\frac{1}{4}\,\zeta'(-1,3/2)\right),
\nonumber\\
&=&\frac{1}{24}\log 2\zeta(-3)-\frac{7}{24}\zeta'(-3)-\frac{1}{24}\log2
-\frac{1}{24}\log
2\zeta(-1)+\frac{1}{24}\zeta'(-1)+\frac{1}{24}\log2,\nonumber\\
&=&-\frac7{24}\zeta_R'(-3)+\frac1{24}\zeta_R'(-1)+\frac1{24}\zeta_R(-3)
\log2-\frac1{24}\zeta_R(-1)\log2,
\end{eqnarray}
where $\zeta_R$ is the Riemann zeta function.

This is the first proper treatment of the zero modes for $S^4$ branes in an 
AdS$_5$ bulk. Similar results have been obtained in lower dimensions by 
Dowker and Apps \cite{DA}.



\end{document}